\newif
\ifconfver
\confvertrue        
\ifconfver
\documentclass[10pt,journal,final,twoside]{IEEEtran}
\else
\documentclass[11pt,draftcls,onecolumn]{IEEEtran}
\fi
\usepackage{placeins,float,bbm,xspace,empheq,fancybox,amsmath,amssymb,graphicx,epstopdf,epsfig,subfigure,syntonly,times,amsthm} 
\usepackage{psfrag,color,bm,array,cite}
\usepackage{url,cite,footnote,xspace,syntonly,algorithm,algorithmic}
\usepackage{verbatim,multirow,slashbox}
\usepackage[T1]{fontenc}
\usepackage{algorithm,cite}
\usepackage{array,booktabs,arydshln}
\usepackage{enumitem,kantlipsum}

\usepackage[dvipsnames]{xcolor}
\usepackage{tikz}
\usetikzlibrary{positioning}

\DeclareMathOperator*{\argmin}{arg\,min}

\allowdisplaybreaks[3]

\newtheorem{remark}{\bfseries Remark}
\input{mysymbol.sty}

\newcommand{\diag}{\mathrm{diag}}

\newcommand{\cblue}{\color{black}{}}

\begin{document}


\title{Topology-aware Graph Neural Networks for \\ Learning Feasible and Adaptive AC-OPF Solutions}
\author{
\IEEEauthorblockN{Shaohui Liu},~\IEEEmembership{Student Member, IEEE}, \IEEEauthorblockN{Chengyang Wu}, and \IEEEauthorblockN{Hao Zhu},~\IEEEmembership{Senior Member, IEEE}

\thanks{\protect\rule{0pt}{3mm} 
    This work has been supported by NSF Awards 1802319 and 2130706. S. Liu, C. Wu, and H. Zhu are with the Department of Electrical \& Computer Engineering, The University of Texas at Austin, 2501 Speedway, Austin, TX 78712, USA; e-mail: {\{shaohui.liu, chengyangwu, haozhu\}{@}utexas.edu}.}
}

\markboth{(revised arXiv preprint)}%
{Liu \MakeLowercase{\textit{et al.}}: Topology-aware Graph Neural Networks for Learning Feasible and Adaptive AC-OPF Solutions}
\renewcommand{\thepage}{}
\maketitle
\pagenumbering{arabic}

%
\begin{abstract}
Solving the optimal power flow (OPF) problem is a fundamental task to ensure the system efficiency and reliability in real-time electricity grid operations. We develop a new topology-informed graph neural network (GNN) approach for predicting the optimal solutions of real-time ac-OPF problem. To incorporate grid topology to the NN model, the proposed GNN-for-OPF framework innovatively exploits the locality property of locational marginal prices and voltage magnitude. Furthermore, we develop a physics-aware (ac-)flow feasibility regularization approach for general OPF learning. The advantages of our proposed designs include reduced model complexity, improved generalizability and feasibility guarantees. By providing the analytical understanding on the graph subspace stability under grid topology contingency, we show the proposed GNN can quickly adapt to varying grid topology by an efficient re-training strategy. {Numerical tests on various test systems of different sizes have validated the prediction accuracy, improved flow feasibility, and topology adaptivity capability of our proposed GNN-based learning framework.}
\end{abstract}


\section{Introduction}
\label{sec:intro}

The optimal power flow (OPF) problem is one of the most fundamental tasks in market operations and power system management. It is instrumental for ensuring high efficiency and security of real-time operations \cite{cain2012history}, particularly under increasingly intermittent and variable energy resources such as renewables 
and flexible demands. 
While there is a growing interest on a machine learning (ML) paradigm for OPF, most OPF learning problems have seldom embraced the physical models of the power grids  nor systematically addressed the critical OPF constraints. 

The accurate ac-OPF problem is known to incur high computation complexity due to its non-linear, non-convex formulation \cite{cain2012history}. With active research on developing direct ac-OPF solvers, growing interest has emerged recently on ML for OPF by obtaining neural network (NN) based prediction models using extensive off-line training; to name a few,  \cite{singh2021learning,gupta2021dnn,baker2019learning,zamzam2020learning,nellikkath2021physics,li2022acpflearning,chatzos2021spatial} for ac-OPF and \cite{pan2019deepopf,zhao2021ensuring,zhang2021convex} for dc-OPF.
Nonetheless, almost all existing work relies on the fully connected NNs (FCNNs) that are agnostic to the power grid topology. As a result, these FCNNs need to be completely re-trained whenever the grid topology and other operation conditions change in daily operations \cite{kassakian2011future}. This lack of topology adaptivity severely affects their adoption by grid operators due to the computation concern. In addition,  the feasibility issue of those OPF learning solutions is very important, especially for the network-wide line limit constraints. For example, a feasible domain technique was developed in \cite{zamzam2020learning,zhao2021ensuring}, {while KKT conditions for ac/dc-OPF were used for training regularization \cite{nellikkath2021physics,zhang2021convex}.} While the first approach could affect the OPF solution optimality, the second one tends to add a large number of regularization terms, all of which require the design of weight coefficients (hyper-parameters). Therefore, it is of great importance to develop a physics-informed OPF learning  framework that can adapt to fast-varying grid topology while simplifying the process of ensuring OPF feasibility.  



The goal of this paper is to leverage the graph neural networks (GNNs) by incorporating the grid topology into a physics-informed OPF learning framework {that extends our earlier work \cite{liu2021graph}}. 
{\cblue When the nodal features exhibit a graph-based \textit{locality property} or topology dependence, the GNN architecture is known to efficiently incorporate the underlying graph embedding; see e.g., \cite{kipf2016semi,garg2020generalization}.} As a special case of NNs, GNNs work for graph learning by aggregating, or filtering the features from neighboring nodes only, thus significantly reducing number of parameters \cite{gama2020graphs}. {\cblue GNNs have been recently used for power system learning tasks such as fault localization in distribution networks \cite{chen2020gcnfault,li2021physics}. While recent work has proposed to use GNNs  for OPF learning \cite{owerko2019optimal,owerko2022unsup}, 
the GNN output labels therein are the nodal power injections which are not topology dependent, critically affecting the prediction performance.} We  put forth an innovative idea of predicting ac-OPF outputs that are topology-dependent, namely the locational marginal prices (LMPs) and voltage magnitudes. Both are recognized to strongly exhibit locality property, due to the power flow (PF) coupling and OPF duality analysis. {\cblue Therefore, our proposed GNN model can effectively utilize the sparse graph embedding underlying these OPF outputs to greatly simplify the model complexity compared with FCNNs}, and thus attain better generalizability during fast-varying operations. The GNN model can be also used for other OPF learning tasks such as line congestion classification.

We further consider two crucial ML-for-OPF enhancements: ac-feasibility and topology adaptivity {\cblue for the proposed GNN model}. To attain the ac-feasibility regularization (FR), we propose to represent the line apparent power flow using the two aforementioned OPF outputs and additional latent variables based on the accurate PF linearization. In addition, the feasibility of other OPF outputs is directly enforced on the prediction outputs. The proposed FR  approach works with fully ac-OPF, which can mitigate data over-fitting issue with improved training time and line flow limit guarantees. As for the topology adaptivity, we advocate that the topology-informed GNN is extremely friendly in transferring to a new topology with a slight modification of the pre-trained graph filter. In addition to this adjustment, a re-training step using new data samples generated for a post-contingency topology can quickly update the GNN models of high prediction accuracy. We analytically support this GNN adaptivity performance by investigating the perturbation bounds on the graph-based subspace, in which the OPF outputs lie. To our best knowledge, the proposed GNN model is the first of its kind in transforming OPF learning to be topology-aware and adaptive. The main contributions of our work are three-fold, as summarized here.
\begin{enumerate}[wide, labelwidth=0pt, labelindent=0pt]
    \item We put forth a GNN-based OPF learning framework as inspired by the topology dependence of certain ac-OPF variables. We demonstrate the GNN advantages including lower model complexity and better generalizability thanks to the sparse grid topology.
    \item We develop a new ac-FR approach for GNN and general NN training using the predicted outputs of prices and voltages, that can reduce line (apparent) power flow violation. This FR approach is fully ac-OPF based and greatly improves the training time and feasibility guarantees of predicted solutions.
    \item We enable the topology adaptivity of our proposed GNN by considering varying grid topology and the stability of the underlying graph subspace. An efficient re-training strategy has been developed to allow fast topology transfer at reduced samples and training time.
\end{enumerate}

The rest of this paper is organized as follows. The (ac-) OPF is formulated in Sec.~\ref{sec:ps}, along with the OPF output analysis including LMPs and voltages. Sec.~\ref{sec:gnn} puts forth the topology-aware GNN-for-OPF learning framework, by designing the loss function and discussing the connection to line congestion classification task. The ac-FR approach is discusses in Sec.~\ref{sec:feas_reg}, while the topology adaptivity feature and the stability of graph filter under topology perturbation in Sec.~\ref{sec:topo}. Numerical results are presented in Sec.~\ref{sec:test}  to demonstrate the prediction and computation performance of the proposed GNN-for-OPF learning at reduced model complexity and improved feasibility, while confirming its topology adaptivity for real-time OPF learning. Sec.~\ref{sec:conclusion} concludes this paper.

\section{Real-Time ac-OPF}
\label{sec:ps}

The real-time optimal power flow (OPF) problem \cite{cain2012history} aims to determine the incremental adjustment to the day-ahead dispatch by obtaining the most economic dispatching signals while accounting for transmission constraints. This work focuses on nonlinear ac-OPF for a power network represented by $\mathcal{G}(\mathcal{V},\mathcal{E})$. The node set $\mathcal{V}$ consists of $N$ nodes, while the edge set $\mathcal{E}\in \mathcal{V}\times \mathcal{V}$ includes transmission lines and transformers. 
Let $\bbp, ~\bbq \in \mathbb{R}^N$ collect the nodal active and reactive power injections, respectively; and similarly for the complex voltage $\bbv \in \mathbb{C}^N$. Given the network admittance (Y-bus) matrix $\bbY\in \mathbb{C}^{N\times N}$, 
the ac-OPF problem can be formulated as
\begin{subequations}
\label{eq:acopf}
\begin{align}
 \min_{\bbp,\bbq,\bbv}~~  & \textstyle\sum_{i=1}^N c_i(p_i) \\
\mbox{s.t.}\;~ &\bbp + \mathrm{j}\bbq = \diag(\bbv)(\bbY\bbv)^{*}, \label{eq:acopf_equal}\\
&\underline{\bbV} \leq |\bbv| \leq \bbarbbV,\label{eq:acopf_v}\\
&\underline{\bbp} \leq \bbp \leq \bbarbbp,~\underline{\bbq} \leq \bbq \leq \bbarbbq, \label{eq:acopf_p}\\
& s_{ij}(\bbv) \leq \bbars_{ij} \,, \,\;\; \forall(i,j) \in \mathcal{E}. \label{eq:acopf_f}
\end{align}
\end{subequations}
where $c_i (\cdot)$ is a convex (typically quadratic or piece-wise linear) cost function for capturing flexible active power injection. Both generation and demand can be flexible, with negative injections indicating the latter. {\cblue The complex power flow equations in \eqref{eq:acopf_equal} follow from the Kirchhoff’s laws}, while the inequality constraints list the operational limits of voltage magnitudes \eqref{eq:acopf_v}, nodal power injections \eqref{eq:acopf_p}, as well as line thermal limits \eqref{eq:acopf_f}, respectively. Unlike power limits, line limits and also voltage constraints depend on system-wide inputs, and thus are difficult to satisfy. {As discussed later, the former will be used as an important regularization term for ensuring OPF feasibility, while the latter can be directly enforced on the prediction outputs.}

Although NN-based models have been recently developed for OPF learning and real-time prediction tasks \cite{baker2019learning,zamzam2020learning}, almost all existing approaches have not addressed model complexity or simplification issue for improving the generalizability. We contend that a graph learning framework is extremely attractive for this purpose, by embedding the underlying grid topology into the NN model. 
To this end, we first consider the input features available for each instance of ac-OPF problem in \eqref{eq:acopf}, which should include uncontrollable components in nodal power injections from fixed demand or renewables. In addition, the cost function $c_i(\cdot)$ depends on the submitted offers/bids, thus varying for each OPF instance as well. Hence, for each node $i\in \ccalV$ the inputs include  $\bbx_i \triangleq [ \bbarp_i, \underline{p}_i, \bbarq_i, \underline{q}_i, \bbc_i ] \in \mathbb{R}^{d}$, with $\bbc_i$ denoting the $(d-4)$ parameters used for defining the nodal cost function. 
For example, a quadratic cost is given by the quadratic and linear coefficients, while piece-wise linear one by the change points and gradients of each linear part. 

To simplify the NN models, the key is to identify OPF outputs that are topology-dependent or exhibit locality property \cite{jia2013impact,ramakrishna2021grid}. To this end, we put forth the optimal solution of voltage magnitudes $|\bbv^*|$ and locational marginal prices (LMPs) $\bbpi^*$ as the outputs to predict ($*$ denoting the optimal values). {\cblue Both are important OPF outputs for grid operations as they are used to determine the voltage set-point and price signal, respectively \cite[Chs.~6\&12]{glover2012power}.} Second, to demonstrate the LMPs' relation to system topology, consider the linearized dc-OPF for solving $\bbp$ only, given by
\begin{subequations}
\label{eq:dcopf}
\begin{align}
\min_{\underline{\bbp} \leq \bbp \leq \bbarbbp} ~~\textstyle & \textstyle \sum_{i=1}^N c_i(p_i) \\
\mbox{s.t.}\;~ &\mathbf{1}^\top \bbp = 0,  \label{eq:dcopf_equal}\\
&  -\bbarbbf  \leq \bbS\bbp \leq {\bbarbbf} \label{eq:dcopf_f}
\end{align}
\end{subequations}
where the injection shift factor (ISF) matrix $\bbS$ forms the active power flows $\bbf = \bbS \bbp$. 
By introducing dual variables $\lambda$, and $[\underbar{\bbmu}$; $\bar{\bbmu}]$  for constraints \eqref{eq:dcopf_equal} and \eqref{eq:dcopf_f}, respectively, the nodal LMP vector becomes 
$\bbpi^* \triangleq \lambda^* \cdot \mathbf{1} -\bbS^\top (\bar{\bbmu}^* - \underbar{\bbmu}^*)$
based on the optimal dual variables. The shadow price $(\bar{\bbmu}^* - \underbar{\bbmu}^*)$ indicates the congested lines; i.e., $\bbarmu_\ell^*(\underline{\mu}_\ell^*) \neq0 $ if and only if line $\ell$ reaches the respective limit. Typically, only a few transmission lines are  congested \cite{price2011reduced}, which leads to the \textit{locality} property of LMPs. {Specifically,  $\bbS$ is formed by topology-based matrices, namely the graph incidence matrix  $\bbA$ and a diagonal matrix with line reactance in $\bbX = \diag\{x_{ij}\}$, as well as the resultant B-bus matrix $\bbB = \bbA^\top \bbX^{-1} \bbA$ as a weighted graph Laplacian. Note that both $\bbA$ and $\bbB$ are reduced from the original matrices by eliminating a reference bus to ensure full-rankness.}  
 
Given the compact singular value decomposition (SVD) $\bbA^\top \bbX^{-\frac{1}{2}} = \bbU\bbSigma\bbV^\top$ and thus eigen-decomposition $\bbB = \bbU\bbSigma^2\bbU^\top$, we can write it as   
\begin{align}
\bbS^\top =  \bbB^{-1} \bbA^\top \bbX^{-1} = \bbU \bbSigma^{-1} \bbV^\top \bbX^{-\frac{1}{2}}. \label{eq:matS}
\end{align}
Thus, the LMP $\bbpi^*$ is generated by the eigen-space $\bbU$ of the graph Laplacian, making it ideal for a graph learning task. 
{\cblue Even though this LMP analysis is limited to the simple dc-OPF, the topology dependence also holds for the ac-OPF case under the strong influence of network congestion on LMP \cite{garcia2019non}. The ac-OPF could be complicated by nonlinear power flow and real-reactive power coupling, further inspiring an OPF learning approach to utilize high prediction capability of NN models.}

Similarly for voltage magnitude, its topology dependence is well known in power flow analysis as exemplified by the fast decoupled power flow (FDPF) model \cite[Ch.~3]{gomez2018electric}. By eliminating the reference bus, we can approximately represent the voltage deviation as
\begin{align}
 \Delta |\bbv| \approxeq -\bbB^{-1}\Delta\bbq
\end{align}
{where matrix $\bbB$ is very close to the reduced B-bus matrix in \eqref{eq:matS}, making the voltage solution $|\bbv^*|$ perfect for graph learning, as well.  For both  $\bbpi^*$ and $|\bbv^*|$, one can view $\bbB$ as the underlying graph shift operator (GSO) \cite{ramakrishna2021grid} that  generates these two OPF output signals, motivating us to pursue a GNN-based OPF learning framework. }  


\section{GNN for OPF Learning}
\label{sec:gnn}

NN-based models have been widely used to learn OPF solution \cite{misra2018learning,deka2019learning,chen2020learning,baker2019learning,pan2019deepopf,zamzam2020learning,zhang2021convex}. The general OPF learning frameworks aim to learn a parametric function $f(\bbX^0;\bbphi) \rightarrow \{\bbpi^*, |\bbv^*|\}$, {\cblue where the input to the first layer, $\bbX^0 \in \mathbb R^{N\times d}$, contains the aforementioned nodal features $\{\bbx_i\}$ as the rows.} Almost all existing OPF learning approaches use the fully-connected NN (FCNN), {\cblue where for each layer $(t+1)$ we have}
\begin{align}
    \mathbf{X}^{t+1}=\sigma(\mathbf{W}^t \mathbf{X}^t+\mathbf{b}^t),~~\forall t = 0,\ldots, T-1
    \label{eq:fcnn_layer}
\end{align}
where $\bbW^t$ and $\bbb^t$ are parameters to learn, and $\sigma(\cdot)$ a nonlinear activation function like ReLU. 
Albeit useful for various \textit{end-to-end} learning tasks, the FCNN models incur significant scalability issue for large power systems. 

\begin{figure}[tb!]
     \centering
         \includegraphics[width=0.8\linewidth]{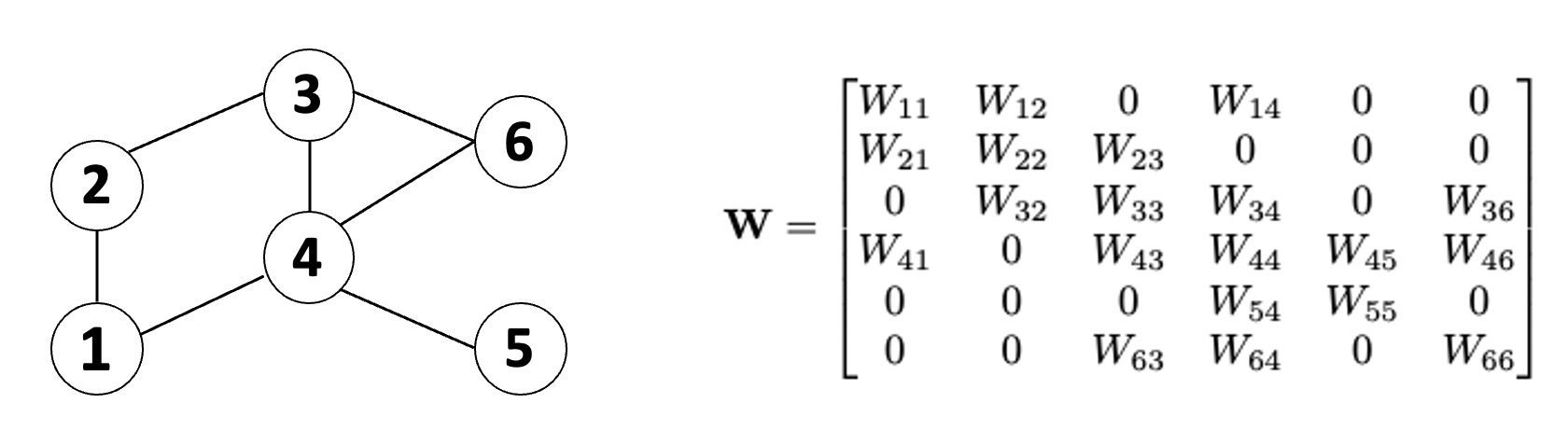}
    \caption{An example 6-bus system and the sparse graph filter $\bbW$. }
    \vspace*{-15pt}
    \label{fig:gidnn}
\end{figure}

Inspired by the graph signal viewpoint on $\{\bbpi^*, |\bbv^*|\}$, we propose to systematically reduce the prediction model complexity by leveraging the GNN models \cite{isufi2020edgenets,ma2020deep,kipf2016semi}. As a special case of NNs, GNNs take the input features $\{\bbx_i\}$ defined over $i\in\ccalV$, with each layer aggregating only the features from  neighboring nodes. In this sense, GNN is ideal for predicting output labels having topology dependency and locality property. To define the GNN layers, consider again the input feature matrix $\bbX^0$, {\cblue and each layer $(t+1)$ now becomes:} 
\begin{align} 
\bbX^{t+1} = \sigma \left( \bbW^t \bbX^{t} \bbH^t + \bbb^t \right),~~\forall t = 0,\ldots, T-1 \label{eq:gnn}
\end{align}
where each graph filters $\bbW^t \in \mathbb R^{N\times N}$ and each feature filter $\bbH^t \in \mathbb R^{d^t \times d^{t+1}}$ constitute as  parameter matrices. Note that $\{\bbH^t\}$ do not increase with the system size $N$. The key of GNNs lies in using graph filters $\{\bbW^t\}$ to aggregate nodal features within a neighborhood only, as illustrated by Fig.~\ref{fig:gidnn}. Specifically, $\{\bbW^t\}$ build upon the graph sparse structure with the location of non-zero entries depending on $\ccalE$, while their values being learned through the training process. Unlike many existing GNN models that fix graph filters with (weighted) graph Laplacian or adjacency matrix, we include $\{\bbW^t\}$ as part of GNN parameters for better prediction accuracy and thus constructing the model \eqref{eq:gnn} becomes a bi-linear problem as discussed in \cite{isufi2020edgenets}.
The output signals of each layer exhibits locality property over the graph, as the filter $\bbW^t$ only aggregates input from the neighboring nodes. The GNN architecture can significantly reduce the number of parameters per layer. As the average node degree of real-world power grids is around 2 or 3 \cite{birchfield2016grid}, we have the following.


\begin{assumption}
\label{as:E}
The edges are very sparse, and the number of edges $|\ccalE| \sim \ccalO(|\ccalV|) = \ccalO(N)$. 
\end{assumption}

\begin{proposition}[GNN complexity order]
Under (AS\ref{as:E}) and by defining $D = \max_t\{d_t\}$, the number of parameters for each bi-linear GNN layer in \eqref{eq:gnn} is $\ccalO(N + D^2)$.
\label{prop:complexity}
\end{proposition}
\vspace*{-2mm}
This complexity order result follows easily from checking the number of nonzero entries in $\bbW^t$ and $\bbH^t$ in \eqref{eq:gnn}. Among the two, only the filter $\bbW^t$ depends on the number of edges, which scales linearly with $N$ thanks to (AS\ref{as:E}). In contrast with $\ccalO(N^2D^2)$ parameters in each FCNN layer, the GNN model greatly improves the scalability with $N$. In addition to reducing computation complexity, the low-complexity GNN layers are instrumental for attaining better generalization performance by mitigating data overfitting and numerical instability issues, based on the statistical learning principles \cite[Ch.~5]{GNNBook2022}.
For example, GNNs have been popularly used for semi-supervised learning tasks with partial observations \cite{feng2020graph,zhang2018_link}. While our work  focuses on full input availability, our numerical tests later on have demonstrated the effectiveness of GNN models in predicting auxiliary power dispatch outputs, thanks to its improved generalizability.

For the GNN loss function, one can compare the predicted $\{\hhatbbpi, |\hhatbbv|\} = f(\bbX^0;\bbphi)$ with actual OPF solutions to form {\cblue 
\begin{align}
        \ccalL \left( \bbphi \right) \!:=\! \gamma_\pi\|{\bbpi^*} - \hhatbbpi \|^2_2 + \gamma_v\big\|{|\bbv^*|} - |\hhatbbv| \big\|^2_2, 
      \label{eq:general_loss}
\end{align}
with the  hyper-parameters $\gamma_{\pi,v}$ as weight coefficients. The hyper-parameters can be selected based on the different variability in predicting $\bbpi^*$ and $|\bbv^*|$, as determined by the NN solvers. In addition to the typical Euclidean norm $\ell_2$, one could use other error norms such as $\ell_1$ error for robustness \cite[Chp.~1]{ben2009robust} or the infinity norm $\ell_\infty$ to penalize the largest deviation  \cite[Appx.~A2]{boyd2004convex}. Our numerical experience suggests that adding an $\ell_\infty$ error norm for $\bbpi^*$ prediction could be useful due to the higher variability of LMP values than voltages.} The design of loss function will be further expanded in the ensuring section to consider the ac-OPF feasibility issue.  Before that, we remark on the applicability of GNN-based models for line congestion prediction here.

\begin{remark}[{\cblue GNNs for Line Prediction}] 
\label{rmk:congestion}
To accelerate numerical OPF solvers, existing work \cite{misra2018learning,deka2019learning} has advocated to classify the binding line/generation constraints using FCNN. Likewise, our proposed GNN models are suitable for this task because line congestion is also topology-dependent. Specifically, the dc power flow model $\bbf = \bbS\bbp$ implies that line flow relates to nodal injection based on the ISF matrix $\bbS$, which depends on the graph eigen-space as discussed in Sec.~\ref{sec:ps}. Therefore, the computation efficiency and generalizability can be similarly achieved by the GNN-based line congestion prediction. For this classification task, the cross-entropy loss function \cite[Ch.~11]{hastie2009elements} will be used. Our numerical experience further suggests to add a fully-connected final layer to match with ISF $\bbS$ which is a full matrix. Notice that number of congested lines is much smaller compared to the total lines \cite{2017congestion}, making it a difficult rare event prediction task. Accordingly, we will focus on predicting the \textit{active lines} that have high likelihood of congestion. {\cblue In addition, this GNN-based line prediction approach could be used for identifying lines to switch for optimal transmission systems, for which FCNN-based learning has been developed in \cite{oren2019switchinglearn}. }
\end{remark}

\section{AC-Feasibility Regularization}
\label{sec:feas_reg}


Regularization can greatly enhance the performance of NN models by mitigating data over-fitting and improving the training speed. Similarly for OPF learning, regularization has been introduced to e.g., improve the constraint satisfaction \cite{zhao2021ensuring}, or approach the first-order optimality \cite{nellikkath2021physics,zhang2021convex}. Nonetheless, the flow constraint  in \eqref{eq:acopf_f} is by and large the most critical feasibility condition of ac-OPF, motivating us to develop a new ac-feasibility regularization (FR) approach. 


{The key of this FR approach lies in generating the line apparent power $s_{ij}$ from the GNN outputs, as the ac power flow admits that 
\begin{align}
{s}_{ij} = |v_i| |I_{ij}| = \Big| |v_i|e^{j\theta_i} - |v_j|e^{j\theta_j} \Big| |v_i||Y_{ij}| 
\label{eq:s_ij}
\end{align}
where $Y_{ij}$ is the admittance for line $(i,j)$ from the Y-bus matrix, while $\{\theta_i\}$ are the bus voltage phase angles. } Using the LMP vector, the latter can be obtained by solving the nodal injection $\hhatbbp$ based on the optimality conditions. This way, our GNN models can generate both $\hhatbbp$ and $\hhatbbtheta$ from $\hhatbbpi$ as latent variables. A feed-forward architecture is possible for ac-FR, as

\tikzset{every picture/.style={line width=0.75pt}} 

\begin{tikzpicture}[x=0.75pt,y=0.75pt,yscale=-1,xscale=1]

\draw    (162.8,123.3) -- (179.25,141.32) ;
\draw [shift={(180.6,142.8)}, rotate = 227.61] [color={rgb, 255:red, 0; green, 0; blue, 0 }  ][line width=0.75]    (10.93,-3.29) .. controls (6.95,-1.4) and (3.31,-0.3) .. (0,0) .. controls (3.31,0.3) and (6.95,1.4) .. (10.93,3.29)   ;
\draw    (162.8,123.3) -- (178.51,104.73) ;
\draw [shift={(179.8,103.2)}, rotate = 130.22] [color={rgb, 255:red, 0; green, 0; blue, 0 }  ][line width=0.75]    (10.93,-3.29) .. controls (6.95,-1.4) and (3.31,-0.3) .. (0,0) .. controls (3.31,0.3) and (6.95,1.4) .. (10.93,3.29)   ;
\draw    (196.6,105.25) -- (221.5,105.25) ;
\draw [shift={(223.5,105.25)}, rotate = 180] [color={rgb, 255:red, 0; green, 0; blue, 0 }  ][line width=0.75]    (10.93,-3.29) .. controls (6.95,-1.4) and (3.31,-0.3) .. (0,0) .. controls (3.31,0.3) and (6.95,1.4) .. (10.93,3.29)   ;
\draw    (246.6,105.03) -- (276.67,104.98) ;
\draw [shift={(278.67,104.98)}, rotate = 179.91] [color={rgb, 255:red, 0; green, 0; blue, 0 }  ][line width=0.75]    (10.93,-3.29) .. controls (6.95,-1.4) and (3.31,-0.3) .. (0,0) .. controls (3.31,0.3) and (6.95,1.4) .. (10.93,3.29)   ;
\draw    (200,144.54) -- (316.18,143.3) ;
\draw   (302.9,103.48) -- (316.18,103.48) -- (316.18,143.3) ;
\draw    (316.18,124.89) -- (341.1,124.44) ;
\draw [shift={(343.1,124.4)}, rotate = 178.78] [color={rgb, 255:red, 0; green, 0; blue, 0 }  ][line width=0.75]    (10.93,-3.29) .. controls (6.95,-1.4) and (3.31,-0.3) .. (0,0) .. controls (3.31,0.3) and (6.95,1.4) .. (10.93,3.29)   ;
\draw    (109.5,123.25) -- (163,123.2) ;

\draw (94.5,121.5) node  [font=\large]  {$\bbX^{0}$};
\draw (141,110.5) node  [font=\footnotesize]  {$f\left({\displaystyle \bbX^{0}\mathbf{,\bbphi }}\right)$};
\draw (188.8,101.3) node  [font=\fontsize{1.29em}{1.55em}\selectfont]  {$\hat{\mathbf{\bbpi }}$};
\draw (189.1,139.5) node  [font=\fontsize{1.29em}{1.55em}\selectfont]  {$|\hat{\mathbf{v}}|$};
\draw (206.6,94) node  [font=\scriptsize]  {\eqref{eq:pg_opt_step}};
\draw (236.11,100.51) node  [font=\large]  {$\hat{\mathbf{p}}$};
\draw (258.03,94.39) node  [font=\scriptsize]  {\eqref{eq:fdpf}};
\draw (291.51,101.07) node  [font=\fontsize{1.29em}{1.55em}\selectfont]  {$\hat{\mathbf{\bbtheta }}$};
\draw (331,112.06) node  [font=\scriptsize]  {\eqref{eq:s_ij}};
\draw (348.07,111.54) node [anchor=north west][inner sep=0.75pt]  [font=\large]  {$\hat{s}_{ij}$};
\end{tikzpicture}
recalling that $f(\bbX^0;\bbphi)$ represents the GNN model of trainable parameter $\bbphi$ according to \eqref{eq:gnn}.

{The implicit KKT optimality condition \cite[Ch.~5]{boyd2004convex} allows to determine the injection $\bbp$ from the LMP, as for each node $i$
\begin{align}
\hat{p}_i = \argmin_{\underline{p}_{i} \leq p_i \leq \bbarp_{i}}~~ &c_i(p_i) - \hat{\pi}_i p_i ~, \label{eq:pi2pg}
\end{align}
which is basically the economic interpretation for OPF. 
For simplicity, consider a quadratic injection cost as $ c_i(p_i) = a_i p_i^2 + b_i p_i $ with $a_i > 0$, and thus the unique optimum becomes 
\begin{align}
\hat{p}_i = 
\begin{cases}
\underline{p}_{i}, &\mathrm{if}~ \frac{ \hat{\pi}_i - b_i}{2a_i} \leq \underline{p}_{i}, \\
\bbarp_{i}, &\mathrm{if}~  \frac{ \hat{\pi}_i - b_i}{2a_i} \geq \bbarp_{i},  \\
\frac{ \hat{\pi}_i - b_i}{2a_i}, &\mathrm{otherwise}.
\end{cases} 
\label{eq:pg_opt}
\end{align}
Similarly for (piece-wise) linear or other costs, the optimal $\hhatp_i$ is determined based on both local derivative $c'_i({p}_i)$ and LMP $\hhatpi_i$. In case of non-unique solutions, the power balance condition is sometimes needed. The nodal relation in \eqref{eq:pg_opt} can be stacked into a vector form, as 
\begin{align}
\hhatbbp &= \mathbb{P}_{[\underline{\bbp},~ \bbarbbp]}\left[\frac{ \hhatbbpi - \bbb}{2\bba}\right]
\label{eq:pg_opt_step}
\end{align}
where $\mathbb{P}[\cdot]$ projects the input to injection limits, while the fractional form represents entry-wise division. 

We use $\hhatbbp$ to further approximate the voltage angle $\hhatbbtheta$  via power flow linearization. Specifically, based on the \textit{decoupled power flow} model \cite[Ch.~6]{glover2012power}, one can linearize around a nominal operating point $\{\bbtheta_o, \bbp_o\}$ to obtain 
\begin{align}
\hat{\bbtheta} \approxeq \bbtheta_{o} + {\bbJ}_{p\theta}^{-1} (\hhatbbp-\bbp_o)
\label{eq:fdpf}
\end{align}
where $\bbJ_{p\theta}$ stands for  the sub-matrix of the power flow Jacobian for the $\bbp$-$\bbtheta$ mapping. Note that the inverse of $\bbJ_{p\theta}$ exists by eliminating the reference bus. {\cblue 
As nodal voltage magnitudes during normal operations are very close to flat voltage, one could simplify \eqref{eq:fdpf} using the dc power flow  $\hat{\bbtheta} \approxeq \bbB^{-1} \hhatbbp$, which turns out effective in our numerical tests.} By and large, the active power only based angle approximation is quite accurate for large-scale transmission systems, as discussed in \cite{dc_revisit2009} and references therein. 

Obtaining $\hat{\bbs}$ is now possible by incorporating the predicted $\hat{\bbtheta} $ and $|\hhatbbv|$ into \eqref{eq:s_ij}, which allows to introduce a regularization term on the total violation level of line flow limit $\bbarbbs$, as given by 
\begin{align}
\mathcal{R}(\bbphi)  := \left\|\mathbb P_{[\mathbf 0, \infty ]}[\hhatbbs - \bbarbbs]\right\|_1
\label{eq:s_infeas}
\end{align}
where $\ell_1$ norm is used to sum up the violation of every line in both directions. {\cblue The loss function in \eqref{eq:general_loss} is updated by
\begin{align}
        \ccalL' ( \bbphi ) :&= \ccalL ( \bbphi ) + \gamma  \mathcal{R}(\bbphi)   \label{eq:loss}
\end{align}
with the FR hyper-parameter $\gamma$.} {\cblue In addition, one could add the prediction error for the latent output, i.e., the nodal injection $\bbp$ in \eqref{eq:pg_opt_step}. Since $\hhatbbp$ is formed from the predicted price $\hhatbbpi$ through a nonlinear projection step, and thus minimizing the error on $\hhatbbpi$ can indirectly reduce the active power prediction error as well. Furthermore, $\hhatbbp$ is actually always feasible. 
}
Here, our ac-FR approach will focus on the line flow FR in \eqref{eq:loss}, due to the difficulty in maintaining the network flow feasibility.  

Before summarizing the feed-forward architecture, we point out the dc-counterpart for our FR approach, given by:}
    \begin{align*}
        \bbX^0 \xrightarrow{\text{$f(\bbX^0;\bbphi)$}} \hhatbbpi
        \xrightarrow{\text{\eqref{eq:pg_opt_step}}} \hhatbbp \xrightarrow{\text{$\bbS\hhatbbp$}} {\hhatbbf} 
\end{align*}
Note that the line active power $\bbf$ can be directly generated by the ISF matrix $\bbS$, making the dc-FR term in \eqref{eq:s_infeas} much easier to form. Our numerical tests will start with the dc-OPF comparisons to showcase the effectiveness of FR, followed up by ac-OPF comparisons for actual performance evaluation. 

\begin{proposition}[Feasibility of GNN predictions] \label{prop:ac-FR}
Thanks to the proposed generation of latent variables in \eqref{eq:pg_opt_step}--\eqref{eq:fdpf}, the ac-FR based OPF learning is a fully feed-forward architecture to minimize \eqref{eq:loss}. Even with the FR term $\mathcal R (\cdot)$, the GNN training is still efficient using \texttt{autograd} and \texttt{backpropagation}. In addition, the feasibility of both predicted $|\hat{\bbv}|$ and ${\hhatbbp}$ can be strictly enforced via projecting onto their respective limits. 
\end{proposition}
{Note that the voltage prediction turns out pretty accurate as that of injection, making it easy to maintain the voltage feasibility in our numerical tests.} Finally, we discuss how to accelerate FR based ac-OPF training in practice.

\begin{remark}[Accelerating FR-based OPF learning] The projection step for producing the latent variables in \eqref{eq:pg_opt_step} can be of computational concern for (G)NN training. As a non-differentiable function, it can affect the convergence speed of gradient-based learning updates. To this end, we can approximate \eqref{eq:pg_opt_step} using the sigmoid activation function $\sigma_s(\cdot)$. Given the input $\bbr:= \frac{ {\hhatbbpi} - \bbb}{2\bba}$, we form the following
\begin{subequations}
\label{eq:pg_opt_sigmoid}
\begin{align}
\bbr'&= \left[ \sigma_s(\underline{\bbp} - \bbr) \right] \cdot (\underline{\bbp} - \bbr) + \bbr,\\
{\hhatbbp} &= \left[ \sigma_s(\bbr' - \bar{\bbp}) \right] \cdot (\bbr' - \bar{\bbp}) + \bar{\bbp}
\end{align}
\end{subequations}
where each $\sigma_s$ soft-thresholds the input $\bbr$ at respective lower/upper limit, yielding a smooth, differentiable mapping. Similarly, one could replace the projection in  \eqref{eq:s_infeas} by the sigmoid as well for smoothness. These modifications can improve the computation efficiency of the FR-based OPF learning with minimal impact on prediction accuracy. 
 \end{remark}



\section{Grid Topology Adaptivity of GNN}
\label{sec:topo}

{\cblue Going beyond scalability and feasibility, it is truly important to enhance the applicability of the proposed GNN-for-OPF learning framework by considering grid  \textit{topology adaptivity}.} 
Almost all existing OPF learning solutions are limited to a fixed topology, and cannot be directly transferred in case of contingency. Nonetheless, the status of lines or transformers is known to change due to contingency or scheduled switching \cite{kassakian2011future,zhou2021}. While this may not be an issue for direct OPF solvers, any variation of topology, or generally operating conditions, require to re-train the NNs after generating new samples. Both sample generation and training lead to concerns over the computation efficiency and real-time adaptivity. To this end, we advocate the proposed GNN models can gracefully address these concerns by analyzing the topology adaptivity performance.  Motivated by OPF security against line outages, we are particularly interested in this type of contingency, while the resultant analysis may be similarly extended to component failures such as generators or loads. 
\begin{assumption}[Topology contingency]
{We consider the contingency of line $k$ outage with no multiple concurrent failures},  and the post-contingency network stays connected. 
\label{as:connectivity}
\end{assumption}
The single-line outage is assumed for simplicity of the analysis, and Sec.~\ref{sec:test} will consider multiple-line outages in practical systems. For topology adaptivity, the key idea is to analyze the perturbation of OPF outputs under (AS\ref{as:connectivity}). Recall that the OPF outputs are generated by the eigen-space of the reduced B-bus $\bbB$ or its inverse, as discussed in Sec.~\ref{sec:ps}. First, consider the perturbation on these two matrices due to (AS\ref{as:connectivity}). Note that $\bbB$ could be written as the aggregation of rank-1 components, as $\bbB = \sum_{\ell \in \mathcal{E}} \frac{1}{x_\ell} \bba_\ell\bba_\ell^\top$ with each vector $\bba_\ell := \bbe_i - \bbe_j$ representing line $\ell$ between buses $i$ and $j$. If line $k$ is outaged, the reduced B-bus matrix becomes ${\bbB}' = \bbB - \frac{1}{x_k} \bba_k\bba_k^\top$ with the inverse 
\begin{align}
({\bbB}')^{-1} = \bbB^{-1} + \bbDelta_k =  \bbB^{-1}+\frac{\bbB^{-1}\bba_k\bba_k^\top\bbB^{-1}}{x_k - \bba_k^\top\bbB^{-1}\bba_k}~.
\label{eq:laplacian_deviation}
\end{align}
Similar to the perturbation of $\bbB$, that of  its inverse under a single line outage is also of rank-one. {For the rest of this paper, we denote the perturbation term in \eqref{eq:laplacian_deviation} as $\bbDelta_k$, {and refer the pre-contingency system by the original system.}

Accordingly, consider the perturbation on the eigen-space of $\bbB^{-1} = \bbU \bbLambda \bbU^\top$, namely the span $span\{\bbU\}$. Let $\bbU'$ concatenate the eigenvectors for $(\bbB')^{-1}=\bbU'\bbLambda'(\bbU')^\top$. We can quantify the difference between the two  {linear} (sub)spaces based on the principal angles between their orthonormal bases \cite[Ch.~13]{gohberg2006invariant}:
\begin{align}
d(span\{\bbU\},span\{\bbU'\}) \triangleq \|\sin \bbTheta \|_F
\label{eq:space_dist}
\end{align}
with $\sin \bbTheta = \diag\left( \sin\theta_1,\cdots,\sin\theta_{N-1} \right)$. 
Note that $\theta_i\triangleq\theta(\bbu_i,\bbu_i')$ refers to the angle between the two unitary  vectors $\bbu_i$ and $\bbu_i'$, with zero angle indicating the same vectors. {Other matrix norms can be used here such as the  Euclidean $\ell_2$ norm.} Interestingly, this distance can be bounded by analyzing the spectrum, or eigenvalues, of $\bbB^{-1}$, related to matrix perturbation theory and invariant subspace concept \cite{gohberg2006invariant}. {Our empirical studies have suggested that some leading eigenvalues of $\bbB^{-1}$ would dominate the subspace for generating $\bbpi$ and $|\bbv|$, which correspond to the smallest eigenvalues of $\bbB$.} If these leading eigenvalues are separable from the remaining ones, we can bound the distance between the two leading subspaces, as established in the following proposition. {Detailed proof is given in Appendix~\ref{app}.}

\begin{proposition}[Bounded subspace difference.] \label{prop:dk_bound} Under (AS\ref{as:connectivity}), both  $\bbB^{-1}$ and $(\bbB')^{-1}=\bbB^{-1}+\bbDelta_k$ are positive definite matrices. {\cblue Let $\{\lambda_i\}_{i=1}^{N-1}$ represent the positive eigenvalues of $\bbB^{-1}$ in a non-increasing order}; e.g., $\lambda_1 \geq \cdots \geq\lambda_{N-1}>0$. {Consider the first $s$ eigenvalues with the minimum separations given by \begin{align*}
			\delta\!\triangleq \! \min_{1\leq i \leq s-1}\!\!(\lambda_{i}-\lambda_{i+1})~\textrm{and}~\delta'\!\triangleq \!\min_{1\leq i \leq s-1}\!(\frac{1}{\lambda_{i+1}}-\frac{1}{\lambda_{i}}).
\end{align*} 
Thus, one can bound the difference between the leading subspace  $span(\bbU_s) \triangleq span(\left[ \bbu_1,\cdots,\bbu_{s} \right])$ with its counterpart $span(\bbU_s')$ as}:
\begin{align}
d(span\{\bbU_s\},span\{\bbU_s'\}) \leq \min \left( \frac{\|\bbDelta_k\|_F}{\delta},\frac{2}{x_k\cdot \delta'}  \right).
\label{eq:subspace_bound_f}
\end{align}
\end{proposition}


The first fractional bound in Proposition~\ref{prop:dk_bound} directly follows from the Davis-Kahan (D-K) Theorem to bound two linear spaces \cite{davis1970rotation,cape2019two}, while the second one is obtained by considering the rank-one perturbation on $\bbB$ itself. 
Note that if the distance function in \eqref{eq:space_dist} is defined based on the $\ell_2$-norm, it is possible to update \eqref{eq:subspace_bound_f} by replacing the bounds in \eqref{prop:dk_bound} by $\min \{ (2\|\bbDelta_k\|_2)/{\delta}, {4}/({x_k\delta'})\}$.

Proposition~\ref{prop:dk_bound} establishes that the subspace that OPF outputs $\bbpi^*$ and $|\bbv^*|$ lie in would remain stable under minor topology perturbations in (AS\ref{as:connectivity}). This is thanks to the small perturbation on the inverse matrix $\bbB^{-1}$ as given by \eqref{eq:laplacian_deviation}. As discussed in Sec.~\ref{sec:ps}, the congestion pattern is typically sparse, which can be further assumed to be stable  under (AS\ref{as:connectivity}). Accordingly, those OPF outputs are likely to stay stable as well, according to the dc-OPF model for generating $\bbpi^*$ in Sec.~\ref{sec:ps}. 
	
To transfer the GNN model trained under the original system to perturbed topology scenarios, we can directly adjust the pre-trained graph filters $\{\bbW^t\}$ in \eqref{eq:gnn} by simply eliminating the entries corresponding to the outaged lines. This direct adjustment step does not require to modify the filters $\{\bbH^t\}$ or any GNN training.
Interestingly, it can likely attain good prediction performance for some line-outage scenarios, as demonstrated by our numerical experience. 
{\cblue To improve the accuracy, the pre-trained model can be used as the warm start for re-training the GNN model with the FR by incorporating additional data samples generated for the new topology. The FR term will require additional changes as detailed in the following remark. 
This re-training step has been shown extremely efficient in terms of convergence rate. To sum up, our proposed GNN architecture can enable fast transfer learning under topology contingency for the OPF learning problem.

\begin{remark}[Feasibility for Topology Adaptive GNNs]
\label{rmk:fr_adaptivity}
As both ac-FR and dc-FR terms depend on the topology, we need to use the perturbed topology when incorporating the predicted $\hhatbbp$ and $|\hhatbbv|$. Specifically, the power-angel sensitivity $\bbJ_{p\theta}$ in \eqref{eq:fdpf} for ac-FR and the ISF matrix $\bbS$ in dc-FR should be updated by the perturbed topology. In addition, for both FR terms one needs to select the set of active lines only. These updates do not affect the GNN model structure for the re-training step. 
\end{remark}

}



\section{Numerical Results}
\label{sec:test}

\begin{figure}[!t]
    \centering
    \vspace*{-5mm}
    \includegraphics[width=90mm]{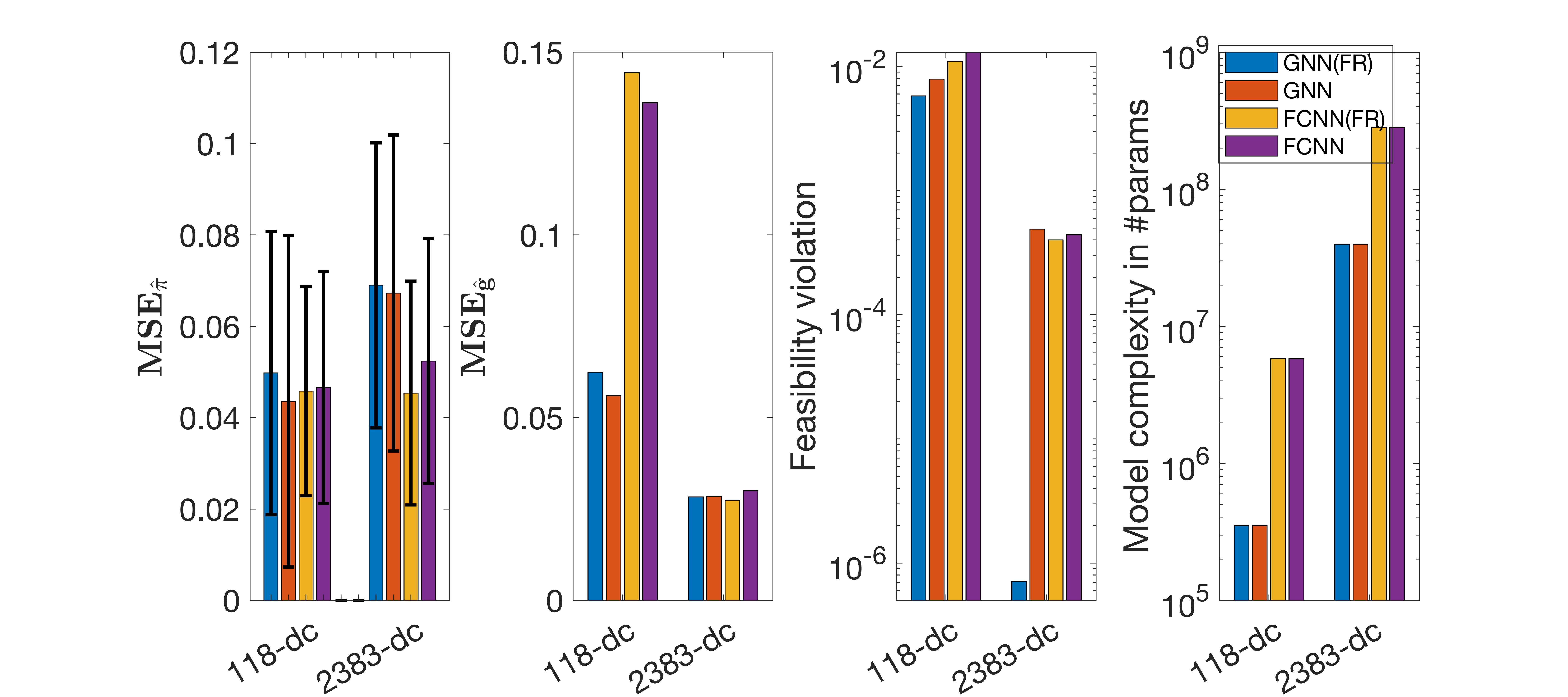}
    \caption{Comparisons of the proposed GNN with FCNN on two dc-OPF test cases (from left to right): i) normalized MSE of predicting $\bbpi^*$ with the its STD; ii) normalized MSE of predicting generator $\bbg^*$; iii) { dc-feasibility violation rate}; and iv) the total number of NN parameters.} 
    \label{fig:perf_stat_dc}
\end{figure}

\begin{figure*}[!t]
    \centering
    \vspace*{-5mm}
    \includegraphics[width=180mm]{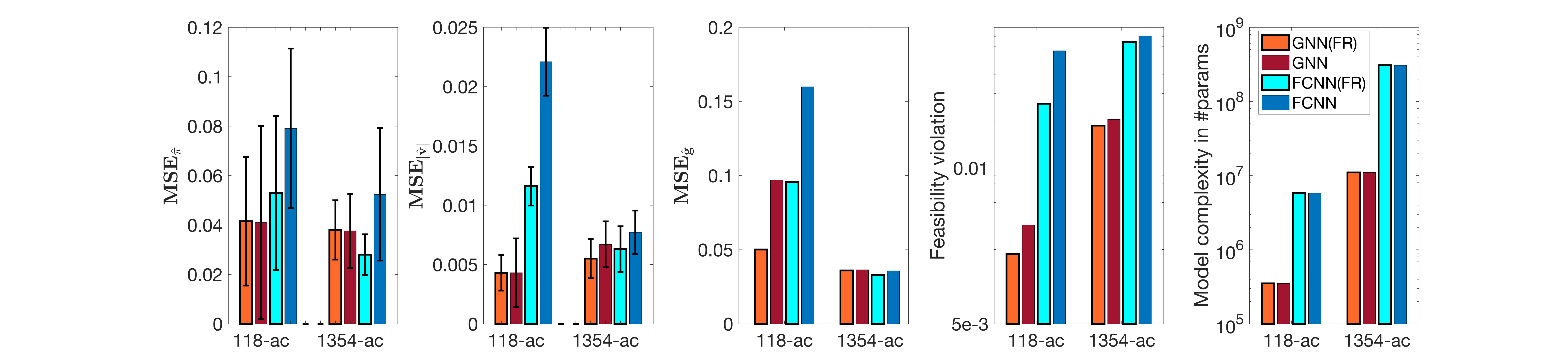}
    \caption{Comparisons of the proposed GNN with FCNN on two ac-OPF test cases (from left to right): i)-ii) normalized MSE of respectively predicting $\bbpi^*$ and $|\bbv^*|$ with its STD; iii) normalized MSE of predicting generator $\bbg^*$; iv) ac-feasibility violation rate; and v) the total number of NN parameters.}
    \vspace*{-5mm}
    \label{fig:perf_stat_ac}
\end{figure*}

\begin{table*}[!ht]
    \centering
    \caption{Comparisons of the proposed GNN with FCNN on ac/dc-OPF test cases in terms of normalized MSE with its STD, feasibility violation rate, and total number of parameters. }
    \begin{tabular}{lllllllll}
     \hline \specialrule{1pt}{0pt}{0pt}
        \textbf{Case} & \textbf{Model} & $\mathrm{\mathbf{MSE}}_{\hat{\mathbf{\pi}}}$ & $\mathrm{\mathbf{STD}}_{\hat{\mathbf{\pi}}}$ & $\mathrm{\mathbf{MSE}}_{\hat{\mathbf{g}}}$ & \textbf{Feasibility} & \textbf{\#Parameters} & $\mathrm{\mathbf{MSE}}_{|\hat{\mathbf{v}}|}$ & $\mathrm{\mathbf{STD}}_{|\hat{\mathbf{v}}|}$ \\ \hline\hline
        \textbf{118-ac} & GNN+FR & 4.15\% & 2.6e-2 & 5.01\% & 2.8e-3 & 351K & 3.4e-3 & 1.5e-3 \\ 
        ~ & GNN & 4.10\% & 3.9e-2 & 9.70\% & 7.4e-3 & 351K & 4.3e-3 & 4.9e-3 \\ 
        ~ & FCNN+FR & 5.30\% & 3.1e-2 & 9.57\% & 2.6e-2 & 5.8M & 1.2e-2 & 1.6e-3 \\ 
        ~ & FCNN & 7.91\% & 3.3e-2 & 15.98\% & 5.7e-2 & 5.8M & 2.2e-2 & 2.9e-3 \\ \hline
        \textbf{1354-ac} & GNN+FR & 3.80\% & 1.2e-2 & 3.60\% & 1.8e-2 & 11M & 5.5e-3 & 1.6e-3 \\ 
        ~ & GNN & 3.76\% & 1.5e-2 & 3.64\% & 2.1e-2 & 11M & 6.7e-3 & 1.9e-3 \\ 
        ~ & FCNN+FR & 2.80\% & 8.2e-3 & 3.29\% & 6.5e-2 & 308M & 6.3e-3 & 1.9e-3 \\ 
        ~ & FCNN & 5.24\% & 2.7e-2 & 3.57\% & 7.0e-2 & 308M & 7.7e-3 & 1.8e-3 \\ \hline
        \textbf{118-dc} & GNN+FR & 4.98\% & 3.1e-2 & 6.24\% & 5.8e-3 & 302K & ~ & ~ \\ 
        ~ & GNN & 4.36\% & 3.6e-2 & 5.60\% & 7.9e-3 & 302K & ~ & ~ \\ 
        ~ & FCNN+FR & 4.58\% & 2.3e-2 & 14.44\% & 1.1e-2 & 5.8M & ~ & ~ \\ 
        ~ & FCNN & 4.66\% & 2.5e-2 & 13.62\% & 1.5e-2 & 5.8M & ~ & ~ \\ \hline
        \textbf{2383-dc} & GNN+FR & 6.90\% & 3.1e-2 & 2.83\% & 7.1e-7 & 39.7M & ~ & ~ \\ 
        ~ & GNN & 6.73\% & 3.5e-2 & 2.85\% & 4.9e-4 & 39.7M & ~ & ~ \\ 
        ~ & FCNN+FR & 4.54\% & 2.5e-2 & 2.74\% & 4.0e-4 & 284M & ~ & ~ \\ 
        ~ & FCNN & 5.24\% & 2.7e-2 & 3.00\% & 4.4e-4 & 284M & ~ & ~ \\ \hline\specialrule{1pt}{0pt}{0pt}
    \end{tabular}
\vspace*{-5mm}
\label{table:benchmark_all}
\end{table*}


This section presents numerical test results for the proposed GNN-based OPF learning framework on several benchmark systems of different sizes\footnote{The codes for dataset generation and test comparisons are available at: \url{https://github.com/cblueLiu/GNN_OPF_electricity_market}.}. We generate data samples using the MATPOWER ac/dc-OPF solvers \cite{matpower}, on the ieee118-bus, pegase1354-bus, and wp2383-bus systems from the IEEE PES PGLib-OPF library \cite{babaeinejadsarookolaee2019power}. For all test cases, the nodal active and reactive power demands and generator cost coefficients have been randomly perturbed from the respective nominal values. Specifically, the load perturbation ranges from 10-30\% with higher perturbation level for smaller systems, where the optimal OPF solutions are obtained by MATPOWER. {\cblue Specifically, the ac-OPF samples are generated by MATPOWER's built-in primal-dual interior point solver (MIPS), while the dc-OPF ones by invoking the Gurobi convex solver \cite{gurobi}.} In addition, all the samples generated are split into the training/test datasets with a 80\%/20\%  division. All learning algorithms have been implemented with the PyTorch library in Python. To train the GNN models, we initialize the graph filters $\{\bbW^t\}$ by the normalized B-bus matrix, and maintain the sparse structure by using a mask matrix. {\cblue The FCNNs are also trained on the same datasets to compare with the GNN performance. For both models, we used a total of six layers where the number of features per node is respectively $\left\{6,5,10,10,5,5\right\}$ for each layer, along with a final linear output layer. As for the dc-OPF training, the number of features per node is given by $\left\{4,5,10,10,5,5\right\}$.}	
All NN models are trained by the standard ADAM algorithm with the same convergence criteria, using NVIDIA Quadro RTX 5000 for computation acceleration.

\textbf{Prediction and Feasibility Performance:} We compare the proposed GNN models 
with the FCNN ones, both having multiple hidden layers. Both types of NN models with and without the proposed FR term, have been considered for the ac-OPF comparisons on the 118-bus and 1354-bus systems, as well as for dc-OPF on the 118-bus and 2383-bus systems. {\cblue For each system, we have generated $10,000$ samples, similar to other OPF-learning work \cite{zhao2021ensuring,nellikkath2021physics,chatzos2021spatial}.} We compare the test performance in predicting both $\bbpi^*$ and $|\bbv^*|$ (in dc-OPF only $\bbpi^*$) measured by the normalized mean squared error (MSE) with its standard deviation (STD). The attained flow feasibility rate is also included as a performance metric. To compare the model complexity, we list the number of parameters in each of the NN models. The test results for dc-OPF cases are presented in Fig.~\ref{fig:perf_stat_dc}, while those for the ac-OPF in Fig.~\ref{fig:perf_stat_ac}. Detailed  results for all test sets are listed in Table~\ref{table:benchmark_all}.

Numerical results for all cases show that the proposed GNN models can achieve comparable accuracy in predicting both OPF solutions of $\bbpi^*$ and $|\bbv^*|$ with respect to the FCNN ones. The voltage prediction is much more accurate than the price one. Both the proposed GNN and ac-FR show improved accuracy, with GNN plus ac-FR showing the best performance in both cases.} In addition, the accuracy of the predictive generation output $\hat{\bbg}$, as part of the nodal injection $\hhatbbp$ in \eqref{eq:pg_opt_step}, is also considered. Notably, the GNN models have shown much higher accuracy in power prediction over the FCNN ones for the 118-dc case. This is likely because the GNN models can  judiciously mitigate the extreme scenarios of $\bbpi$ prediction errors, e.g., those very high price mismatches in certain buses. Note that GNN models use significantly less parameters compared to FCNN, and the reduction is even more evident for larger systems. In addition, the proposed GNN predictions can reduce the occurrence of flow violations as compared with the FCNN ones. Thus, all test results verify the aforementioned advantages of the proposed GNN models in terms of reduced model complexity and improved generalizability, as discussed in Sec.~\ref{sec:gnn}. 

Moreover, we want to point out the effectiveness of the proposed (ac-)FR approach for each of the NN models. Note that this approach is applicable to both GNN and FCNN models, while its improvements have been generally observed for all cases, with the highest reduction for 2383-dc case. Interestingly, while the OPF output prediction error can slightly increase, the error variance for predicting $\bbpi^*$ tends to decrease in all test cases. Hence, our proposed FR approach can improve the feasibility performance while attaining more consistent price prediction. This is extremely valuable for ac-OPF learning, for which flow feasibility is much more difficult to verify compared with dc-OPF.

\begin{table}[tb!]
	\centering
		\caption{{Comparisons of the proposed GNN with FCNN on two test
			systems for congested line classification task.}}	\label{table5}
	\begin{tabular}{lllll}
		\hline\specialrule{1pt}{0pt}{0pt}
		\textbf{ Case} & \multicolumn{2}{c}{\textbf{118-ac}} & \multicolumn{2}{c}{\textbf{2383-dc}} \\ \hline
		Metric & Recall & F1 score & Recall & F1 score \\
		\hline
		GNN & 98.40\% & 96.10\% & 90.00\% & 81.40\%\\
		FCNN & 97.70\% & 94.60\% & 87.30\% &78.30\%\\
		\hline\specialrule{1pt}{0pt}{0pt}
	\end{tabular}
    \vspace*{-5mm}
\end{table}


{\textbf{Line Congestion Prediction:} We have tested the proposed GNN models on the congestion classification task, as discussed in Remark~\ref{rmk:congestion}. To show the main congestion performance, we have first picked the top 10 congested lines in the system, and listed the resultant classification metrics on those lines in Table~\ref{table5}. Specifically, the recall and F1 score  metrics\footnote{{\cblue The two metrics are defined as: $\text{recall}\triangleq\frac{\text{true~positive}}{\text{true~positive+false~negative}}$, and $F1\triangleq\frac{2\cdot\text{recall}\cdot \text{precision}}{\text{recall+precision}}$ where $\text{precision}\triangleq\frac{\text{true~positive}}{\text{true~positive}+\text{false~positive}}$.}}
ideally should approach 100\%, to demonstrate a good balance between sensitivity and accuracy especially for heavily biased datasets. While GNN and FCNN show comparable performance on the smaller 118-bus case, the improvement by the proposed GNN model is clearly observed in the larger 2383-bus case. Therefore, at reduced complexity GNNs can potentially improve the generalization in predicting congested lines for OPF learning.}

\textbf{Topology Adaptivity under Line Outages:} To verify the topology adaptivity as discussed in Sec.~\ref{sec:topo}, we have further considered the {118-ac} and 2383-dc cases with line outages. {\cblue We have generated $8,000$ samples for each system under the original topology.} While our analytical results focused on single-line outages, this numerical test pursues multiple-line outages which can more strongly support the GNN topology adaptivity. For the 118-ac, at most double-line outages have been considered, while for the 2383-dc at most 4-line outages. We have picked 6-7 specific scenarios of line outage contingencies for each test case, by selecting the lines with high likelihood of being congested. {\cblue For each post-contingency new topology (denoted by \texttt{Sys\#} later on), $4,000$ data samples have been re-generated for the re-training step using the random load perturbations as before.} Note that the number of re-training samples is only half of that of the original training process. 

\begin{figure}[!t]
    \centering
    \hspace*{-5mm}
    \vspace*{-5mm}
    \includegraphics[width=100mm]{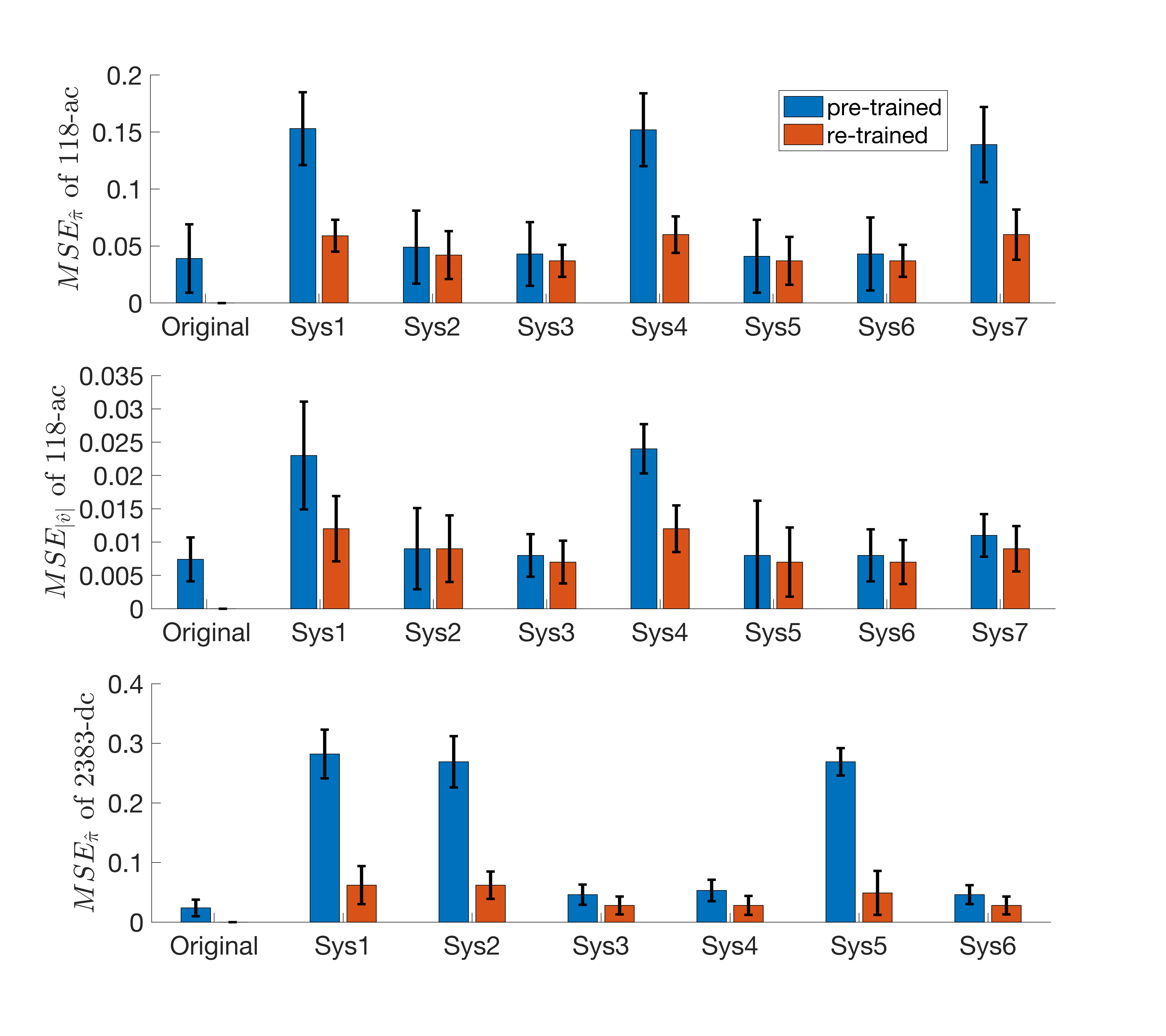}
    \caption{{\cblue Error performance of the pre-trained and re-trained GNN models on each new topology (\texttt{Sys\#}) for the 118-ac (top) and 2383-dc (bottom) systems, in terms of both the normalized MSE with its STD for price $\bbpi^*$ prediction, and $|\bbv^*|$ prediction (ac only).}}
    \vspace*{-5mm}
    \label{fig:118ac2383dc_topo_stat}
\end{figure}

\begin{figure}[t!]
        \centering
        \includegraphics[width=90mm]{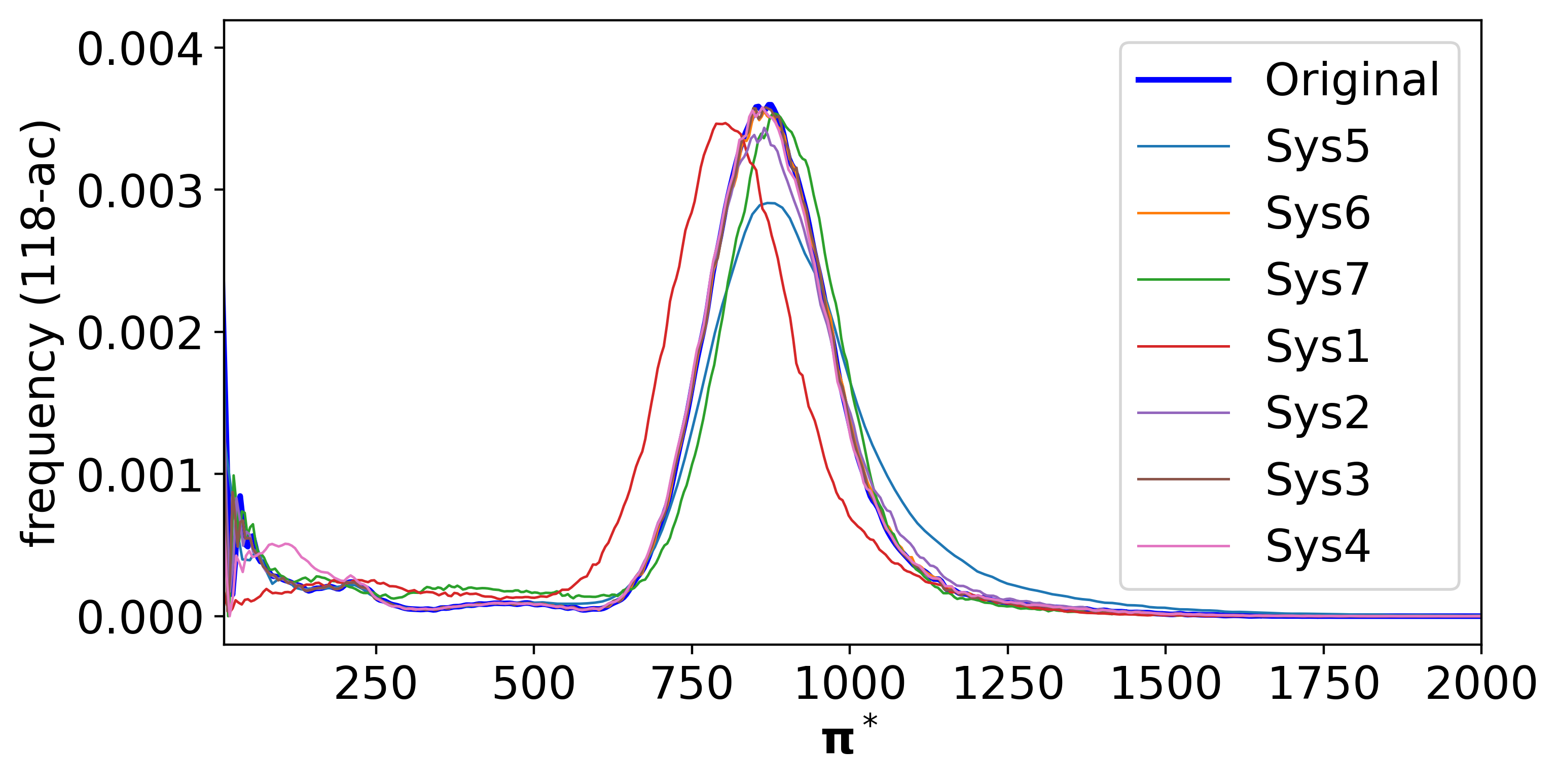}
        \hspace*{-3mm}
        \includegraphics[width=90mm]{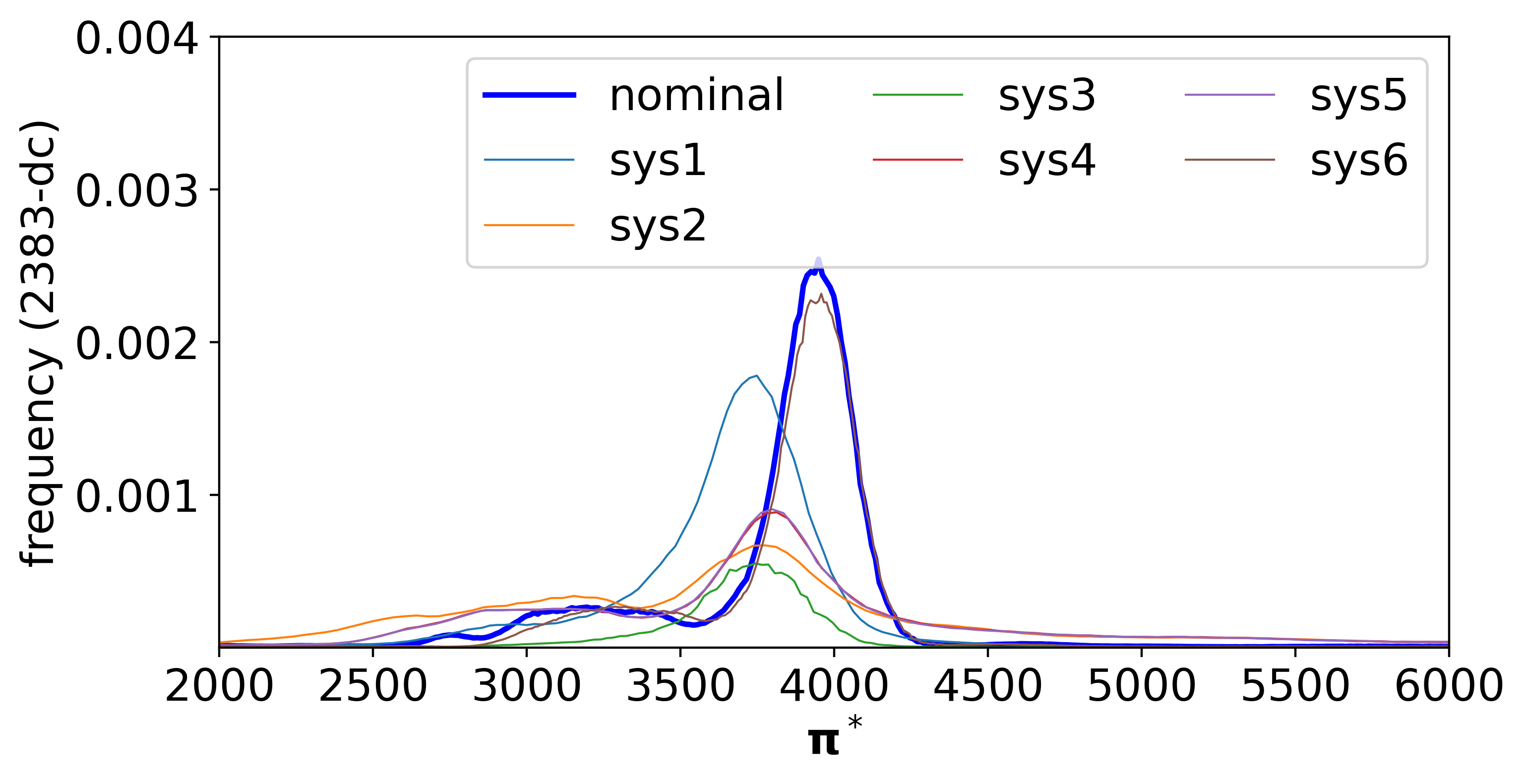}
        \caption{{Histograms of the optimal $\bbpi^*$ for the 118-ac (top) and 2383-dc (bottom) cases with both the original system (bolded blue curve) and each new topology.}
        \label{fig:118n2383_pi_topo_dist}}
\end{figure}

\begin{table}[tb!]
\centering
\caption{ {\cblue Performance statistics of GNN models in both the original system and each new topology for the 118-ac and 2383-dc cases: normalized MSE of predicting $\bbpi^*$ and  $|\bbv^*|$, and their STD for pre-trained and re-trained models (the latter in parentheses), ac/dc feasibility, training time, the subspace and filter perturbations.} }
\begin{tabular}{lllll}
\hline 
\textbf{ 118-ac }& \textbf{Original}   & \texttt{Sys1}   & \texttt{Sys2}    & \texttt{Sys3}    \\ \hline
$\mathrm{\mathbf{MSE}}_{\hat{\mathbf{\pi}}}$ (\%)   & 3.9   & $15.3~(5.9)$ & $4.9~(4.2)$ & $4.3~(3.7)$  \\
$\mathrm{\mathbf{STD}}_{\hat{\mathbf{\pi}}}$ (\%)   & 3.0     & $3.2~(1.4)$ & $3.2~(2.1)$ & $2.8~(1.4)$ \\
$\mathrm{\mathbf{MSE}}_{|\hat{\mathbf{v}}|}$ (\%)   & 0.74   & $2.3~(1.2)$ & $0.9~(0.9)$ & $0.8~(0.7)$  \\
$\mathrm{\mathbf{STD}}_{|\hat{\mathbf{v}}|}$ (\textperthousand)   & 3.3     & $8.1~(4.9)$ & $6.1~(5.0)$ & $3.2~(3.2)$ \\
Feasibility($s$,\%)   & 2.6     & $5.8~(3.2)$ & $4.3~(2.5)$ & $5.6~(2.2)$ \\
Time (s) & 90.2 & 4.9 & 5.1 & 4.8  \\
$d(\cdot,\cdot)$ &  0 & 0.089  &  0.020 & 0.006  \\
$\Delta_\bbH$    & 0    & $0.051$ & $0.040$ & $0.032$ \\\hline 
\textbf{ 118-ac }   & \texttt{Sys4}   & \texttt{Sys5 }  & \texttt{Sys6}    & \texttt{Sys7}  \\  \hline  
$\mathrm{\mathbf{MSE}}_{\hat{\mathbf{\pi}}}$ (\%)   & $15.2~(6.0)$ & $4.1~(3.7)$  & $4.3~(3.7)$ & $13.9~(6.0)$  \\
$\mathrm{\mathbf{STD}}_{\hat{\mathbf{\pi}}}$ (\%)   & $3.2~(1.6)$ & $3.2~(2.1)$ & $3.2~(1.4)$  & $3.3~(2.2)$ \\
$\mathrm{\mathbf{MSE}}_{|\hat{\mathbf{v}}|}$ (\%) & $2.4~(1.2)$ & $0.8~(0.7)$ & $0.8~(0.7)$ & $1.1~(0.9)$  \\
$\mathrm{\mathbf{STD}}_{|\hat{\mathbf{v}}|}$ (\textperthousand) &  $3.7~(3.5)$     & $8.2~(5.2)$ & $3.9~(3.3)$ & $3.2~(3.4)$ \\
Feasibility($s$,\%)   & $4.8~(3.6)$ & $5.3~(2.2)$ & $4.8~(2.7)$ & $4.8~(2.6)$ \\
Time (s) & 4.8 & 5.3 & 6.0 & 6.0  \\
$d(\cdot,\cdot)$  &  0.099 & 0.016  &  0.003 &  0.009 \\
$\Delta_\bbH$   & $0.059$ & $0.033$ & $0.037$ & $0.056$ \\\hline\hline 
\textbf{2383-dc} & \textbf{Original}   & \texttt{Sys1}   & \texttt{Sys2}    & \texttt{Sys3}   \\\hline
$\mathrm{\mathbf{MSE}}_{\hat{\mathbf{\pi}}}$ (\%)   & 2.38   & $28.2~(6.2)$ & $26.9~(6.2)$ & $4.6~(2.8)$ \\
$\mathrm{\mathbf{STD}}_{\hat{\mathbf{\pi}}}$ (\%)   & 1.4     & $4.1~(3.2)$ & $4.3~(2.3)$ & $1.7~(1.5)$ \\ 
Feasibility($f$, \textperthousand)   & 2.6     & $6.6~(3.5)$ & $6.7~(3.5)$ & $7.5~(4.1)$ \\
Time (s) & 126 & 9.1 & 9.2 & 9.0 \\ 
$d(\cdot,\cdot)$ & 0 & 0.012  &  1.5e-3  & 1e-4   \\
$\Delta_\bbH$    & 0    & $0.348$ & $0.217$ & $0.084$ \\ \hline
\textbf{2383-dc} & \textbf{Original}  & \texttt{Sys4}   & \texttt{Sys5 }  & \texttt{Sys6}     \\ \hline
$\mathrm{\mathbf{MSE}}_{\hat{\mathbf{\pi}}}$ (\%)   & 2.38   & $5.3~(2.8)$ & $26.9~(4.9)$ & $4.7~(2.8)$ \\
$\mathrm{\mathbf{STD}}_{\hat{\mathbf{\pi}}}$ (\%)   & 1.4   & $1.8~(1.6)$ & $2.3~(3.7)$ & $1.6~(1.5)$ \\ 
Feasibility($f$, \textperthousand)   & 2.6     & $7.5~(4.1)$ & $6.6~(3.5)$ & $6.5~(4.1)$ \\
Time (s) & 126  & 9.3 & 9.0 & 9.8 \\ 
$d(\cdot,\cdot)$ & 0 &   2e-3  &  0.012  &  1e-4 \\
$\Delta_\bbH$   & 0   & $0.124$ & $0.083$ & $0.015$ \\\hline 
\label{table:118ac2383dc_topo_stat}
\end{tabular}
\vspace*{-5mm}
\end{table}

{Fig.~\ref{fig:118ac2383dc_topo_stat} plots the performance of the pre-trained and re-trained GNN models on each new topology in both systems, in terms of the normalized MSE and its STD in predicting $\bbpi^*$. Detailed results for this performance comparison, including the training time and subspace distance (to be discussed later), are listed Table~\ref{table:118ac2383dc_topo_stat}.} {For each system, the pre-trained GNNs can well adjust to the new topology and produce reasonably prediction accuracy even under the multiple line-outage contingency. }The prediction performance under this direct adjustment slightly degrades in the larger 2383-dc system. Regardless, the re-training process further improves the accuracy which is consistent with that of the original systems (even lower in some cases), using just half of the samples. Table~\ref{table:118ac2383dc_topo_stat} also shows that the re-training process is extremely fast and converges within 10 seconds, as compared to the original training time at dozens of seconds. We have observed that only a handful of epochs, like 5 epochs, are sufficient for the  convergence in most re-training cases. Thus, compared to topology-agnostic FCNN, our proposed GNN models have {demonstrated high efficiency and accuracy in adjusting to a new grid topology even under multiple}, simultaneous line outages.

Furthermore, we investigate the subspace perturbations related to Proposition~\ref{prop:dk_bound} for each new topology. We choose $s=10$ and $s=50$ as the dimensions of subspaces for the 118-bus and 2383-bus systems, respectively. Using principal component analysis (PCA) \cite[Ch.~12]{james2013introduction}, {these dimensions have been determined to ensure that each subspace contains at least 50\% of energy in the original spectrum of $\bbB^{-1}$. {\cblue This way, the two eigenvalue separation parameters in Proposition \ref{prop:dk_bound} are $\delta=0.03$ and $\delta'=0.14$ for the 118-bus system, and $\delta=0.005$ and $\delta'=0.006$ for 2383-bus. Accordingly, the bound in \eqref{eq:subspace_bound_f}  is numerically 28.38 for the 118-bus system, and $1.02e+5$ for 2383-bus, respectively.} {Nonetheless, the actual distances are significantly smaller, as shown by the $d(\cdot,\cdot)$ values in Table~\ref{table:118ac2383dc_topo_stat}. These values imply that the subspace is very stable under the topology perturbation considered, even with multiple line outages. 
{\cblue Furthermore, we have compared the normalization perturbation of feature filters $\{\bbH^t\}$ between the new topology and original one, as given by $\Delta_\bbH :=\frac{1}{T}\sum_{t=0}^{T-1} \frac{\| \bbH_{i}^{t} -  \bbH_{0}^{t} \|_2}{\| \bbH_{0}^{t} \|_2}$ with the subscript indicating system index. Interestingly, the GNN filters are observed to be minimally affected by the topology perturbation, especially for the smaller system. 
}

The stability results are further confirmed by the histograms of the optimal $\bbpi^*$ for each new topology, as plotted in Fig.~\ref{fig:118n2383_pi_topo_dist}. Specifically, the new distributions for the smaller 118-bus system are pretty consistent with each other, with \texttt{Sys4} leading to the largest perturbations in terms of both the subspace and price distribution. As for the 2383-bus system, \texttt{Sys1} gives the highest perturbations in both aspects, as well. Therefore, these numerical results confirm that our proposed subspace perturbation analysis is useful for understanding the impact of topology changes on OPF outputs, explaining why the proposed GNN models exhibit nice topology adaptivity.}




\section{Conclusion and Future Work}
\label{sec:conclusion}

This paper develops a new topology-informed GNN approach for predicting the optimal solutions of real-time ac-OPF problem. We put forth the GNN-based prediction of LMPs and voltage magnitudes, two important OPF outputs, that can capitalize on their topology dependency. The resultant GNN enjoys simplified model structure with significantly reduced number of parameters thanks to a sparse grid topology, and thus achieve good generalization performance. To further enhance the feasibility guarantees of OPF learning, we design an (ac-)feasibility regularization (FR) approach that can effectively reduce the line (apparent) power violation. Going beyond a fixed topology, we investigate the topology adaptivity of GNN models under line contingency, by providing the stability analysis of the graph subspace. Numerical results on various test systems have validated the performance of the proposed GNN models with the FR approach in terms of attaining high prediction accuracy and increasing flow feasibility at significantly fewer parameters. 
The GNN topology adaptivity is also confirmed with high computation efficiency in transferring to a new grid topology.

Several interesting future directions open up for this work. First, the feasibility regularization can be extended to more rigorous risk-based metrics such as conditional-variance-at-risk \cite{lin2022cvar}. 
{\cblue Second, the GNN-based line prediction task could be used to determine line switching decisions for the optimal transmission switching problems; see e.g., \cite{oren2019switchinglearn}.
Last, to consider a multi-stage formulation that incorporates temporal constraints due to ramping and energy storage, the proposed GNN framework could be used to simplify the static PF modeling as done in \cite{kody2022modeling}.} 

\begin{appendices}

	\section{Proof of Proposition~\ref{prop:dk_bound}}\label{app}
	\begin{IEEEproof} First, as the grid topology stays connected under (AS\ref{as:connectivity}), both $\bbB$ and $\bbB'$ as reduced graph Laplacian matrices should be of full rank. Thus, both matrices are positive definite and the corresponding eigenvalues are all positive. 
		
	Both  bounds in \eqref{eq:subspace_bound_f} follow from the Davis-Kahan (D-K) theorem that characterizes subspace perturbation based on the matrix spectrum, as detailed in \cite{cape2019two,yu2015useful}. We have restated the results for the convenience of this proof.
		
	\begin{theorem}[Davis-Kahan Theorem] \label{thm:d-k}
		{Let $\bbM,\bbM'\in\mathbb R^{M\times M}$ be symmetric matrices with $\bbM'$ is a perturbation of $\bbM$.} Their respective eigenvalues are given by $\mu_1\geq \mu_2 \geq \ldots \geq \mu_M$ and $\mu'_1\geq \mu'_2 \geq \ldots \geq \mu'_M$. Let $\bbV_m=[\bbv_1,\cdots,\bbv_m]\in\mathbb R^{M\times m}$ and $\bbV_m'=[\bbv_1',\cdots,\bbv_m']\in\mathbb R^{M\times m}$ be the collections of (orthonormal) eigenvectors corresponding to the first $m$ eigenvalues of $\bbM$ and $\bbM'$, respectively. By defining the minimum separation $\delta_m := \min_{1\leq i \leq m} (\mu_i -\mu_{i+1})$ and $\sin \bbTheta_m := \diag\left( \sin\theta(\bbv_1,\bbv_1'),\cdots,\sin\theta(\bbv_m,\bbv_m') \right)$, we have 
        \begin{align}
        \|\sin \bbTheta_m \|_2 \leq \frac{2\|\bbM-\bbM'\|_2}{\delta_m} ~,
        \label{eq:dk_l2norm}
        \end{align}
        and
        \begin{align}
        \|\sin \bbTheta_m \|_F \leq \frac{\|\bbM-\bbM'\|_F}{\delta_m}.
        \label{eq:dk_fnorm}
        \end{align}
	\end{theorem}

For ease of reference, Theorem~\ref{thm:d-k} is a variation from the original Davis-Kahan $\sin\Theta$ theorem in \cite{davis1970rotation}. The $\ell_2$-norm bound is derived by considering the separation $\delta_m$ of the leading $m$ eigenvalues of $\bbM$ and the corresponding eigenvectors in Theorem 6.9 of \cite{cape2019two}. Similarly, the Frobenius-norm bound can be derived based on Theorem 1 of \cite{yu2015useful}. The rigorous analysis is based on the operator theory and matrix spectrum analysis. For example, in \cite[Sec.~VII.3]{bhatia2013matrix} a generalized version of D-K theorem, namely Theorem VII.3.2 therein, has been developed using the operator theory, while the analysis of Theorems V.3.4 and V.3.6 has been performed based on matrix spectrum in \cite{stewart1990matrix}.
		
For the first bound in Proposition~\ref{prop:dk_bound}, it directly follows from the D-K theorem by considering the perturbation from $\bbM = \bbB^{-1}$ to $\bbM'= (\bbB')^{-1} = \bbB^{-1} + \bbDelta_k$. The fractional term in \eqref{eq:dk_fnorm} of Theorem~\ref{thm:d-k} becomes ${\|\bbDelta_k\|_F}/{\delta}$ using the minimum separation $\delta$.

As for the other bound, we can apply the Davis-Kahan theorem to analyze the perturbation on $\bbB$ to $\bbB'$ under (AS\ref{as:connectivity}). First, note that $\bbB$ and $\bbB^{-1}$ share the same eigenvectors, as $\bbB^{-1} = \bbU\bbLambda\bbU^\top$ leads to  $\bbB = \bbU\bbLambda^{-1}\bbU^\top$. Thus, the matrix $\bbU_s$ also captures the eigenvectors corresponding to the smallest $s$ eigenvalues of $\bbB$, where $\delta'$ defined in Proposition \ref{prop:dk_bound} represents its minimum separation from the other eigenvectors. In addition, recall that $\bba_k := \bbe_i - \bbe_j$ is the vector representing line $k$ between say, buses $i$ and $j$, and thus it has exactly 2 non-zero entries. When line $k$ is outaged under (AS\ref{as:connectivity}), the Frobenius norm of the rank-one perturbation of $\frac{1}{x_k} \bba_k\bba_k^\top$ equals to $\frac{2}{x_k}$. Thus, when $\bbM = \bbB$ is perturbed to $\bbM'= \bbB + \frac{1}{x_k} \bba_k\bba_k^\top$, the bound in \eqref{eq:dk_fnorm} of Theorem~\ref{thm:d-k} becomes $\frac{2}{x_k\cdot\delta'}$ using the corresponding separation $\delta'$.

To sum up, the results in Proposition \ref{prop:dk_bound} is essentially the minimum of two aforementioned bound terms.
%
It is worthy pointing out that the $\ell_2$-norm based results in \eqref{eq:dk_l2norm} can be similarly applied to generate the corresponding bounds for bounding $\|\sin \bbTheta \|_2$, as given by
\begin{align}
\|\sin \bbTheta \|_2 \leq \min \left( \frac{2\|\bbDelta_k\|_2}{\delta},\frac{4}{x_k\cdot \delta'}  \right).
\label{eq:subspace_bound_l2_appendix}
\end{align}
This completes the proof of Proposition \ref{prop:dk_bound} and its $\ell_2$-norm extension as stated in Sec.~\ref{sec:topo}. 

\end{IEEEproof}
\end{appendices}

%
\bibliographystyle{IEEEtran}
\itemsep2pt
\bibliography{IEEEabrv,ref}

\end{document}